\documentclass[12pt]{article}

\usepackage{epsfig}
\usepackage{cite}
\usepackage{amsmath, amssymb, amsfonts}
\usepackage{color}
\usepackage{latexsym}
\usepackage{graphicx}

\def\be{\begin{equation}}
\def\ee{\end{equation}}
\def\ba{\begin{eqnarray}}
\def\ea{\end{eqnarray}}

\def\bdm{\begin{displaymath}}
\def\edm{\end{displaymath}}
\def\la{~\mbox{\raisebox{-.6ex}{$\stackrel{<}{\sim}$}}~}
\def\ga{~\mbox{\raisebox{-.6ex}{$\stackrel{>}{\sim}$}}~}
\def\bq{\begin{quote}}
\def\eq{\end{quote}}

 at 10truept

\newcommand{\de}{\partial}
\newcommand{\rmd}{\mathrm{d}}

\renewcommand{\[}{\left[}
\renewcommand{\]}{\right]}
\renewcommand{\(}{\left(}
\renewcommand{\)}{\right)}

\newcommand{\al}{\alpha}

\newcommand{\om}{\omega}

\newcommand{\bea}{\begin{eqnarray}}
\newcommand{\eea}{\end{eqnarray}}

\newcommand{\bi}{\begin{itemize}}
\newcommand{\ei}{\end{itemize}}

\newcommand{\beq}{\begin{equation}}
\newcommand{\eeq}{\end{equation}}
\newcommand{\beqa}{\begin{eqnarray}}
\newcommand{\eeqa}{\end{eqnarray}}

\def\la{~\mbox{\raisebox{-.6ex}{$\stackrel{<}{\sim}$}}~}
\def\ga{~\mbox{\raisebox{-.6ex}{$\stackrel{>}{\sim}$}}~}

\def\ltap{\ \raise.3ex\hbox{$<$\kern-.75em\lower1ex\hbox{$\sim$}}\ }
\def\gtap{\ \raise.3ex\hbox{$>$\kern-.75em\lower1ex\hbox{$\sim$}}\ }
\def\gl{\ \raise.5ex\hbox{$>$}\kern-.8em\lower.5ex\hbox{$<$}\ }
\def\roughly#1{\raise.3ex\hbox{$#1$\kern-.75em\lower1ex\hbox{$\sim$}}}

\begin{document}

\thispagestyle{empty}
\begin{flushright}
December 2019 \\
\end{flushright}
\vspace*{.75cm}
\begin{center}
  
{\Large \bf On Black Hole Echoes}

\vspace*{1.5cm} {\large Guido D'Amico$^{a, }$\footnote{\tt
damico.guido@gmail.com} and Nemanja Kaloper$^{b, }$\footnote{\tt
kaloper@physics.ucdavis.edu} }\\
\vspace{.3cm} {\em $^a$Dipartimento di SMFI dell' Universit\`a di Parma and INFN, Gruppo Collegato di Parma, Italy}\\
\vspace{.3cm}
{\em $^b$Department of Physics, University of
California, Davis, CA 95616, USA}\\

\vspace{1.5cm} ABSTRACT
\end{center}
We consider a very simple model for gravitational wave echoes from black hole merger ringdowns which may arise from local Lorentz symmetry violations that modify graviton dispersion relations. 
If  the corrections are sufficiently soft so they do not remove the horizon, the reflection of the infalling waves which trigger the echoes is very weak.
As an example, we look at the dispersion relation of a test scalar field corrected by roton-like operators  depending only on spatial momenta, in Gullstrand-Painlev\'e coordinates. The near-horizon regions of a black hole do become reflective, but only very weakly. The resulting ``bounces" of infalling waves can yield repetitive gravity wave emissions but their power is very small.
This implies that to see any echoes from black holes we really need an egregious departure from either standard GR or effective field theory, or both. One possibility to realize such strong echoes is the recently proposed classical
firewalls which replace black hole horizons with material shells surrounding timelike singularities.

\vfill \setcounter{page}{0} \setcounter{footnote}{0}

\vspace{1cm}
\newpage

After much expectation and some trepidation, gravitational waves from black holes have been detected \cite{LIGO} 
ushering an era of gravity wave astronomy. The idea is that gravity waves may become a versatile tool to study 
gravitating structures in the universe; specifically they can be a tool to understand more closely
the structure of their sources. Since black holes are presumed to be among the sources of gravity waves, 
this gives us a chance to learn more about
black holes by studying gravity waves emitted by them. Even very basic questions about black holes are interesting
since now they might be able to be subjected to an experimental test. 
Black holes in standard General Relativity (GR) are perfect absorbers. If a wave is aimed at a black hole,
it will be absorbed, with the cross section given by the black hole horizon area \cite{gibbons}. In a way,
as a wave approaches the near-horizon region, the universality of the horizon geometry suppresses the
reflection of the wave, and instead provides it with an infinitely deep crevasse to slide down. In perturbation theory,
this is seen by the fact that the general-relativistic nonlinearities modify the usual Newtonian centrifugal barrier 
surrounding a gravitating mass and turn it into an infinitely deep potential well. In coordinates which are analytic across the horizon, 
this means that the propagating modes for each helicity are focused by gravity very strongly, with only infalling waves
being a regular mode near the horizon. Technically, the would-be outgoing modes would have unbounded backreaction on the
geometry close to the horizon, and hence having a horizon means that such modes cannot emanate from it. 

This would not happen if geometry were different. Deviations of the geometry from the near horizon geometry of a black hole would generically obstruct perfect absorption and allow a fraction of the wave to reflect. Interestingly, if this were to happen, the reflected wave could end up generating a sequence of wave echoes  \cite{cardoso}, bouncing between the reflective 
region deep inside, where the near horizon geometry is modified, and the interior of the centrifugal potential barrier,
located at distances comparable to the gravitational radius of the source. The reflected wave would trickle out at
regular periods, and if it did, one could obtain some information about the distortion close to where the horizon should
be from correlating the echo period, the reflected amplitudes of different rings, and the frequency dependence of the
observables. Alternatively a failure to observe any echoes would reinforce the universality of 
GR -- and the standard equivalence principle -- near black hole horizons.  

This has prompted a number of 
authors to consider such gravitational wave echoes as a probe of gravity at very high scales, where new phenomena
might arise. Indeed, if there are deviations from standard GR at high scales, the black hole absorptivity might be
imperfect, or even more dramatically, the object might not even be a black hole \cite{cardoso,afshordi,burgess,emparan,foitkleban,kleban,coates}. However unlikely such scenarios might be, it is of interest to consider them given that we really have very few direct tests of gravity in such highly nonlinear regimes. 

In this work we will instead take a more conservative approach and consider breaking of Lorentz symmetry at high energies, which affects only the dispersion relation of the propagating probe wave.
The reason is that even such relatively plain physics could in principle lead to the emergence of gravitational echoes even in the standard black hole geometry. The echoes do arise, but the power they carry is generically very small. The reason is that the black holes in such 
theories {\it are} black: they have regular horizons, with leading-order universal gravity, which therefore makes the corrections only
weakly reflective.

Specifically, the Lorentz violations are already constrained by observations to be quite small at the level of observable phenomena (see, e.g. \cite{glashow,kostelecky,mattingly,ellis,liberati}). This means that any Lorentz violations which one might wish to consider are to be coded by irrelevant operators in the effective field theory (EFT), which are suppressed by some high UV scale. 
If they are to correct dispersion relations that directly affect the propagation of ``free fields" (which still gravitate) on a background, these operators should be quadratic in fields, but involve higher derivatives. In the linear limit (i.e. for fields propagating in weak gravitational fields) such terms are negligible at low energies. 

However, a wave packet falling into a black hole is like a kinetic energy projectile, going in with a steady acceleration and reaching relativistic velocities near the horizon. In a locally Lorentz-invariant theory, if there is nothing but the horizon ahead of it, the wave packet will continue on in, with its image fading away from sight. Essentially, in the tortoise coordinate, the wave packet will behave as a free wave in empty space, which -- after passing through a centrifugal barrier -- goes on all the way. 

If on the other hand Lorentz symmetry violations change the dispersion relation,  the infalling wave  packet will encounter an additional bump -- a well or a barrier -- that shows up far along after the centrifugal hurdle was cleared. 
Just like in ordinary quantum mechanics, a wave packet encountering any such bump will experience a fractional reflection, and a fraction of the energy of the initial wave will bounce off and travel back. The reflected wave will propagate to the centrifugal barrier, and some will get out, while a fraction will reflect back in, triggering a sequence of echoes. 

Note that introducing a Lorentz-violating operator corresponds to favoring a specific set of coordinates.
A problem arises in black hole backgrounds near the horizon, where due to the blueshift the irrelevant operators in generic coordinates -- which depend on a positive power of energy, and so a positive power of a large boost factor -- introduce much worse singularities than one encounters in standard GR coupled to a locally Lorentz-invariant field theory. If permitted, such behavior renders any perturbative description of the wave packet propagation on the background totally unreliable. Note that this is a generic property of Lorentz-violating operators: due to Lorentz breaking, we cannot just go to the locally freely-falling coordinate frame where the theory reduces to
a smooth set of weakly coupled harmonic oscillators which have finite proper stress-energy. When the theory is Lorentz-invariant, we can 
transition to such a system, even near -- and across -- the black hole horizon, when we follow infalling null geodesics, as long as those exist. 
With generic Lorentz breaking, however, infalling geodesics will not be smooth on the horizon, because of the backreaction  of Lorentz-breaking contributions
on the background geometry generated by tidal forces -- i.e., because of too strong focusing in the near horizon limit. Thus any Lorentz breaking
which one is to use as a `perturbation' near the black hole horizon must be rather special in order to remain weak and not disqualify the 
starting background in the first place. This is really nothing new, in fact, and so we will not dwell on further details -- to convince oneself, all
one needs is dimensional analysis and power counting of the boost factors. However we should
be clear about this from the outset. The physical cause of these problems -- the untamed boosts near the black hole -- 
is the same issue which embodies the spirit of black hole no hair theorems and balding dynamics \cite{israel,bekenstein,price}. 

Indeed a similar problem has been encountered in the past, where 
in order to study effects of Lorentz breaking at very short distances (aka ``trans-Planckian" problem) irrelevant 
Lorentz-breaking operators were introduced in Gullstrand-Painlev\'e (GP) coordinates \cite{jacobson,unruh}, 
where such pathological
singularities are precluded by writing them in a coordinate system regular across the horizon\footnote{Any coordinate system which is regular on the horizon would do as well, since the transformations between it and the GP coordinates remain finite. For example we could have used the Eddington-Finkelstein coordinates too.}. Thus these divergences are tamed\footnote{In contrast, \cite{afshordi} add Lorentz-breaking operators in a coordinate system which diverges on the horizon. This feeds the coordinate divergence into the observables, rendering their framework  completely different from a black hole: it is a naked singularity through and through and so completely unreliable.
The claims of echo detection by these works has been questioned in for example \cite{ashton}.}.
We will see that in this case, when we use ingoing GP coordinates as a framework 
to code Lorentz violations, the regularity of the coordinates on the horizon implies that the reflection is small, and the echoes are very quiet. 

In hindsight, this outcome is not surprising. The simple reason why the echoes are quiet is decoupling. The black hole emits the bulk of its gravity waves at frequencies $\simeq 1/r_0$, where $r_0$ is the gravitational radius of the black hole.
If the departures from the perfect black hole configurations of standard GR occur only at short distances,
then in the observed range, the emitted power in echoes cannot be larger than $\sim (\omega_0/k_0)^2 \sim 1/(k_0 r_0)^2$, just like in the studies of trans-Planckian cosmology \cite{brand,shenker}. This scaling is precisely what we find. When the scale $k_0$ controlling the correction is high, since the typical black hole gravity wave frequency is so low, $\omega \simeq 1/r_0 \simeq 10^{-9} {\rm eV}$, the echoes will be invisible. This is much worse than in early universe cosmology, where the frequency can be much higher, $\omega \simeq H \la 10^{13} {\rm GeV}$. To have echoes at the 
observable level, therefore, requires much more dramatic violations of either GR or EFT, or both. 

One might be misled to infer that our conclusion is of limited validity, since -- as we said right up front -- 
it is based on employing a special set of Lorentz-violating
operators: those defined in coordinate charts which are analytic across the horizon. One might object that the effects we see are small because the sources are small. However -- we stress again -- that while
the Lorentz-violating operators we use do seem special -- in that they are introduced in coordinate systems analytic across the horizon -- this is 
necessary in order to keep the whole system under control. The exact same operators which we use could be introduced with
less circumspection, in e.g. Kruskal coordinates, where the echos would then seem to be dramatically larger \cite{afshordi}. However
in this case there are {\it no} modes of this `perturbation' that can ever cross the horizon smoothly - any attempt to pass would
blow the horizon up. Such behavior can be readily divinated from carefully inspecting modes of field theories near the horizon devised
to study Hawking radiation; for examples see \cite{unruhh,pata,cutoffs}. As a result using coordinates which are not analytic
across the horizon as the playground to study effects of Lorentz violation is seriously misguided. The converse outcome,
which we find from employing analytic charts, may seem less exciting but it is reliable, as long as we think the sources of gravity waves are black holes. To argue this, our example is as general as it needs to be. Note that while we have utilized 
Gullstrand-Painleve\'e coordinates in our analysis, we could have used any other coordinate system related to these by a {\it finite} boost
and arbitrary spatial coordinate transformations (as, e.g. ingoing Eddington-Finkelstein). The conclusions would be completely the same.
If the
sources of gravity waves are not black holes, but systems of naked singularities, other approaches may
be applicable. 
We in fact outline one such possibility at the very end of this article, with the understanding that it is not a black 
hole in the usual sense of the word -- since there are no horizons. 

We start with looking at Lorentz violating operators. The Schwarzschild metric in ingoing GP coordinates, in which the time is the proper time of a free-falling observer, is given by
\be
\rmd s^2 = - \rmd t_{GP}^2 + (\rmd r + v(r) \rmd t_{GP})^2 + r^2 \rmd \Omega^2 \, ,
\label{metric}
\ee
where the function $v(r) = - \sqrt{r_0 / r}$, $r_0 = 2 G M$ being the position of the horizon.
The orthonormal tetrad adapted to the free-fall frame is given (in the $t_{GP}-r$ subspace) by the vectors $s_{GP}= (1, v(r), 0, 0)$, $u_{GP} = (0, 1, 0, 0)$. Note that on the horizon, $v = -1$ and thus the metric is
$ds^2 \rightarrow -2dt_{GP} dr + dr^2 + r_0^2 d\Omega^2$, which indeed is regular.

We will probe this geometry with a massless scalar field. The propagation of our probe scalar field on this background is described by \cite{jacobson}
\be
S = \frac{1}{2} \int \rmd^4 x \sqrt{-g} \[ - \de_\mu \phi \de^\mu \phi + \frac{\al}{k_0^2} \(u^\mu \de_\mu (u^\nu \de_\nu \phi) \)^2 \] \, ,
\label{action}
\ee
where the Lorentz-violating operator involves a special frame selected by the ingoing radial vector $u$. In flat space this simulates a quartic spatial momentum correction similar to the rotons in a fluid. 
Here, $k_0$ is a high-momentum cutoff and $\alpha$ a dimensionless number which can take either sign.

We remind the reader that here we study the echos generated by scalar waves falling into a static Schwarzschild black hole
and yet we are extracting the information about the gravity waves falling into an astrophysical, spinning, black hole. While this may
seem off, our argument applies because in theories which are locally Lorentz-invariant to leading order, the counting of degrees
of freedom in linearized approximation shows that the system always reduces to a set of Klein-Gordon
equations, one for each helicity \cite{russrep}, and where near the horizon we can neglect spin-dependent corrections as long as the black hole
spin is far from extremal \cite{page}. We treat Lorentz breaking as a background supported correction, and so our precise numerics might
be off, but as long as the corrections are finite, these will only appear as infra-red effects.
Further, real black holes have spin, so one may wonder that the spin significantly modifies our results.
This may be; but as long as the black hole is non-extremal these should be subleading
corrections only.
Even with spin, the echos will stay tiny, as long as the corrections to the near-horizon spacetime happen at very small distances.

It is now straightforward to obtain the field equation for the scalar, by varying (\ref{action}) and writing the result on the background (\ref{metric}). 
After expanding the field into spherical harmonics, $\phi = \sum_{\ell, m} \frac{\Psi(t_{GP}, r)}{r} Y_{\ell m}(\hat{n})$, and performing straightforward albeit tedious algebra, we find the equation of motion
\ba
&&- \ddot{\Psi} + v \( \frac{\dot{\Psi}}{2 r} - 2 \dot{\Psi}' \) + \frac{v^2}{r} \Psi' + (1-v^2) \Psi''
- \[ \frac{\ell (\ell+1)}{r^2} + \frac{r_0}{r^3} \] \Psi + \nonumber \\
&& ~~~~~~~~~~~~~~~~~~~~~~~~~~~~~~~~~~~~~~~~~~~~~~~~~~~ 
+ \frac{\al}{k_0^2} \[ \de_r^4 \Psi + \frac{2}{r^2} \Psi'' - \frac{4}{r^3} \Psi' + \frac{4}{r^4} \Psi \] = 0 \, .
\label{eomGP}
\ea
The terms which are $\alpha$-independent are readily recognized as the field equation
for a massless scalar propagating on the Schwarzschild geometry, $\Box \phi =0$, whereas 
the $\alpha$-dependent terms are the Lorentz-violating corrections projected onto the 
radial vector $u$, which picks the special frame. Because of this choice, the corrections depend only on the
radial derivatives, correcting the dispersion relation of $\phi$. 

Note that in the limit where we approach the horizon, $v = -1$, the coefficient of the term $\propto \Psi''$ vanishes.
This means that the horizon is a regular-singular point of (\ref{eomGP}), and hence only one of the solutions
of (\ref{eomGP}) can be regular on it. That solution on the horizon obeys the limiting form of (\ref{eomGP}) which,
in the absence of the $\alpha$-dependent terms, involves only first derivatives in the radial direction. It 
is easy to check that when $\alpha=0$ and near the horizon $v=-1$, where we can ignore the effects of the 
centrifugal barrier, the equation for monochromatic waves $\Psi = e^{-i\omega t_{GP}} \zeta$ degenerates to 
$\zeta' = - i \omega \zeta/2$ describing precisely the infalling modes as the only regular solutions. 

The next step is to determine how the correction $\propto \alpha$ changes this conclusion. However, continuing with the analysis in GP coordinates turns out to be cumbersome. 
Transitioning to the standard Schwarzschild coordinates in the Kruskal gauge helps tidy up the calculations 
of the wave reflection, since the equation can be readily rewritten as a Schr\"odinger equation and the Lorentz-violating terms can be treated as a perturbation. 
The important aspect of this coordinate change is to {\it preserve} the special selection of
the Lorentz-violating operator $\propto \alpha$, selected in the GP coordinates to avoid introducing uncontrollable divergences on the horizon. This means that in the Schwarzschild coordinates, the vector $u_{GP} = (0,1,0,0)$ will not  have such a simple component form, but will be transformed by the time coordinate change,
\be
\rmd t_S = \rmd t_{GP} + \frac{v(r)}{1-r_0/r} dr \, , 
\label{time}
\ee
while the radial coordinate remains the same.  Note that this means that simply taking Eq. (\ref{eomGP}) and replacing $t_{GP}$ with $t_S$ is {\it incorrect}, because this would miss the transformation of $u$ induced by
the radially dependent time-shift (\ref{time}). Instead, we will have to reevaluate the $\alpha$-dependent terms
in the new coordinates, using the transformed set of  components for $u$. 
Henceforth, we will drop the subscripts $S$, and use $r$ and $t$ for the standard Schwarzschild coordinates. 

Now we turn to transforming the field equation (\ref{eomGP}), using (\ref{time}). Here our strategy will be to first consider the Lorentz-invariant terms, $
\Box \phi = 0$, determine the leading order solution, and then calculate the leading correction to it by an iterative method, where we plug the leading order solution in the $\alpha$-dependent terms and use the result as a source for determining the Lorentz-violating correction to $\phi$. The correction in principle contains both the projections on
the infalling and reflected wave, and so once we compute it we will need to project out the infalling contribution to
finally determine the reflected wave. 

So, using the same expansion in spherical harmonics as before, but with the Schwarzschild time as opposed to
GP time, $\phi = \sum_{\ell, m} \frac{\Psi(t, r)}{r} Y_{\ell m}(\hat{n})$, we find that the 
the Lorentz-invariant part -- i.e., the D'Alembertian -- transforms to 
\be
\Box \phi \rightarrow - \frac{1}{r - {r_0}} \de_t^2 \Psi + \frac{r-r_0}{r^2} \de_r^2 \Psi  + \frac{r_0}{r^3} \de_r \Psi  - \frac{\ell (\ell + 1) r - r_0}{r^4} \Psi \, .
\label{boxsch}
\ee
Since the Schwarzschild time $t$  is a Killing direction, we can now Fourier transform 
$\Psi(t,r) = e^{-i\omega t} \psi(r)$. Substituting this in (\ref{boxsch}) and using the tortoise radial coordinate,
\be
r_* = r + r_0 \ln\left(\frac{r}{r_0} - 1\right) \, ,
\ee
helps organize the radial derivatives. We finally obtain 
\be
\Box \phi \rightarrow \frac{1}{r-r_0} \( \frac{\rmd^2 \psi}{\rmd r_*^2} + \om^2 \psi \) - \frac{\ell (\ell+1) r - r_0}{r^4} \psi \, ,
\label{eq:freeeq}
\ee
which takes the form of a Schr\"odinger equation. 
The potential term has a maximum at $r \sim O(r_0)$, and close to the horizon vanishes exponentially in $r_* \to - \infty $. 
Since we are interested in the dynamics very close to the horizon, we will generally neglect this potential. 
Imposing purely ingoing boundary conditions at the horizon, reflecting the GP analysis above, our zeroth order solution near the horizon is then simply 
\be
\psi_{0} = e^{-i \om r_*} \, .
\label{zeroth}
\ee

To calculate the correction due to the Lorentz-breaking terms, we recalculate the $\alpha$-dependent terms in the new coordinate system, and substitute $\psi_0$ in the terms proportional to $\alpha$. The fact that $\psi_0$ solves the free wave equation then implies that the $\alpha$-dependent terms evaluated on $\psi_0$ source the correction $\psi_1$, and so on. 
After a straightforward but tedious calculation, we obtain the equation  for the correction,  in the near-horizon limit where we neglect the centrifugal barrier
\be
\label{eq:firstorder}
\begin{split}
\( \frac{\rmd^2 \psi_1}{\rmd x_*^2} + f^2 \psi_1 \) = ~ & \frac{\al e^{-i f x_*}}{(k_0 r_0)^2} \frac{\sqrt{x} - 1}{(\sqrt{x} + 1)^3}
\Bigl\{ f^4 x + 3 i f^3 x^{-1/2} - f^2 \frac{7 + 4 \sqrt{x} + 8 x}{4 x^2} + \\
&+  i f \frac{21 + 68 \sqrt{x} + 73 x + 32 x^{3/2}}{8 x^{7/2}} + 4 \frac{(\sqrt{x} + 1)^4}{x^5} \Bigr\} \equiv e^{-i f x_*} \mathcal{S}(x) \, .
\end{split}
\ee
Here we have introduced dimensionless variables $x = r/r_0$, $x_* = r_* / r_0$ and $f = \om r_0$.
As $r \to r_0$, $x \to 1$, and it can be easily checked that the right hand side vanishes as $~(\sqrt{x}-1)$. 
We stress that utilizing the tortoise radial coordinate is more useful than the Schwarzschild radial coordinate, 
in order to understand that the solutions near the horizon remain free of spurionic singularities. The notation in 
(\ref{eq:firstorder}) is cumbersome, since it involves both standard and tortoise radial coordinates, which we
-- as everyone else  -- use merely for the compactness of the notation.

After ignoring the centrifugal barrier, the LHS  of (\ref{eq:firstorder}) is especially simple, which enables us to use the Green's function technique to solve  
for $\psi_1$.  The required  Green's function is just  the harmonic oscillator with appropriately chosen boundary conditions. Since we are not modifying the physics at the horizon, and the correction vanishes there, we impose purely ingoing boundary conditions as $x_* \to - \infty$. However, since we are interested in the reflected wave, for $x_* \to + \infty$ we want both ingoing and outgoing waves. It is straightforward to show that the Green's function satisfying these boundary conditions is
\be
G(x_*, y_*) = c \, e^{-i f (x_* - y_*)}
+ \frac{1}{2 i f} \[ e^{i f (x_*-y_*)} \theta(x_* - y_*) +
e^{-i f (x_*-y_*)} \theta(y_* - x_*) \] \, ,
\ee
where $c$ is a constant, determined by the normalization.

\begin{figure}[h]
  \centering
  \includegraphics[width=0.49\textwidth]{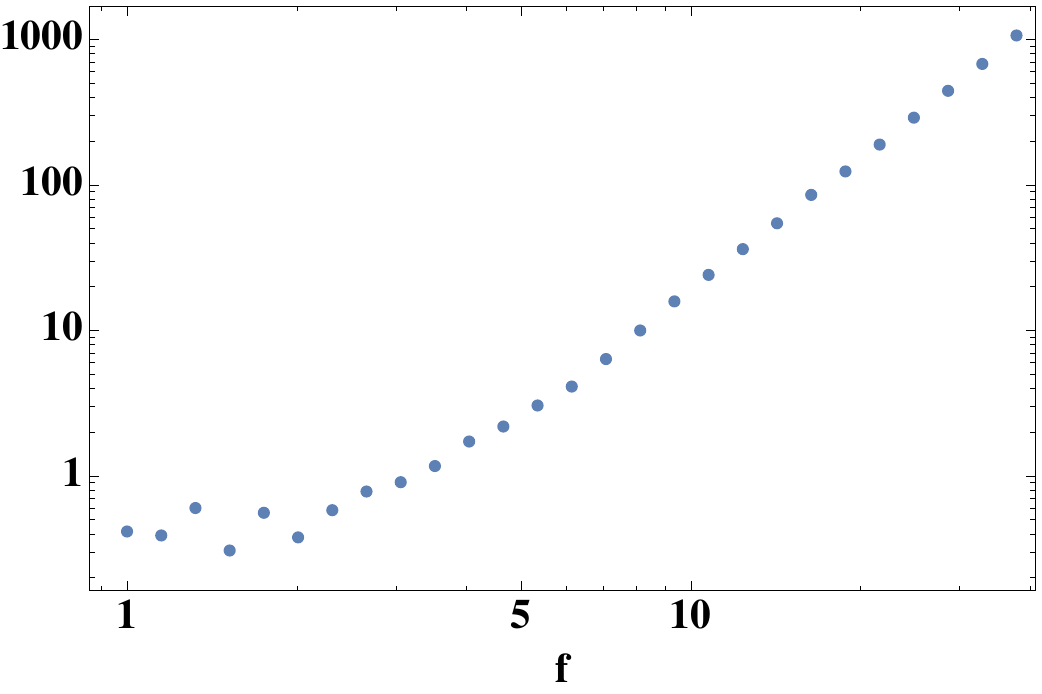}
  \includegraphics[width=0.49\textwidth]{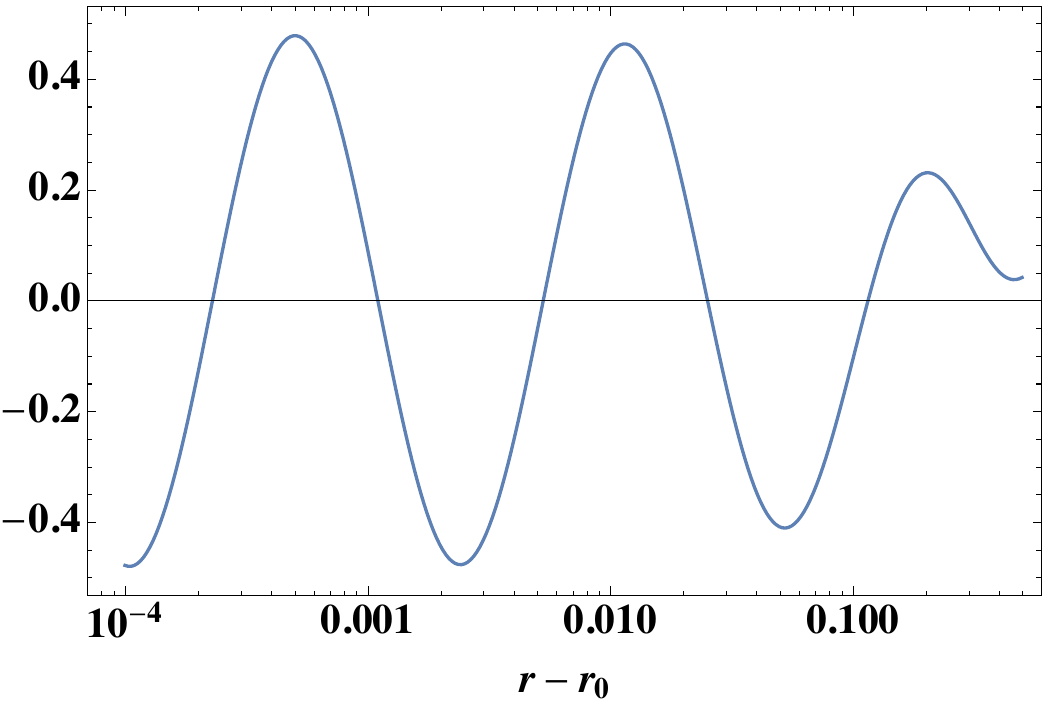}
  \caption{\textbf{Left Panel:} Plot of the numerical evaluation of $k_0^2 r_0^2 |R| / \alpha$, where $R$ is the reflection coefficient, as a function of dimensionless frequency $f$. \textbf{Right Panel:} Plot of the real part of the perturbation $k_0^2 r_0^2 \psi_1(r) / \alpha$, at fixed frequency $f = 2$, as a function of the distance from the horizon $r-r_0$.}
  \label{fig:R}
\end{figure}

The leading order correction to the zeroth order solution is therefore
\beq
\psi_1(x_*) = \int_{-\infty}^{x_*} \rmd y_* \frac{e^{i f (x_* - 2 y_*)}}{2 i f} \mathcal{S}(y(y_*)) + 
\int_{x_*}^{\infty} \rmd y_* \frac{e^{-i f x_*}}{2 i f} \mathcal{S}(y(y_*))
+ c \, e^{-i f x_*} \int_{-\infty}^{\infty} \rmd y_* \mathcal{S}(y(y_*)) \, ,
\label{eq:psi1}
\eeq
where the last term is a purely incoming mode.
A numerical solution for $\psi_1$ is shown in Fig.~\ref{fig:R}.
To identify the amplitude of the reflected wave, we project $\psi_1$ on $\(e^{i f x_*}\)^* = e^{-i f x_*}$. This yields
\be
R = \int_{-\infty}^{\infty} \rmd x_* \, e^{-i f x_*} \psi_1(x_*) = R^{(a)} + R^{(b)} \, ,
\label{intr} 
\ee
where we have
\be
R^{(a)} = \frac{1}{2 i f} \int_{-\infty}^{+\infty} \rmd x_*  \int_{-\infty}^{x_*} \rmd y_*  \,  e^{-2 i f y_* }  \, \mathcal{S}(y(y_*)) \, ,
\label{inta}
\ee
and
\be
R^{(b)} = \frac{1}{2 i f} \int_{-\infty}^{+\infty} \rmd x_*  \, e^{-2 i f x_*} \int_{x_*}^{+\infty} \rmd y_*  \, \mathcal{S}(y(y_*)) \, .
\label{intb} 
\ee
These integrals are cumbersome, but the exact answers aren't necessary to understand the physics. 
We will  therefore only focus on the limits of high and low frequencies. In the limit $\om r_0 \gg 1$, we have
\be
\mathcal{S}(x) = \frac{\al f^4}{(k_0 r_0)^2} \frac{x (\sqrt{x} - 1)}{(\sqrt{x} + 1)} \, .
\label{Shigh} 
\ee
In this case, the integrals for $R$ diverge at large distances: however, this is because we neglected the centrifugal barrier term in eq. (\ref{eq:freeeq}), having focused on the solutions close to the horizon.
Therefore, we can regulate these IR divergences by introducing a cutoff at the far end of the barrier -- ie the photon sphere --  $r \sim \frac{3}{2} r_0$. This is morally equivalent to the IR cutoff in the brick wall model for black hole entropy \cite{thooft}. 

In this case, we find that the reflection coefficient is dominated by $R^{(b)}$, and a numerical evaluation gives the absolute value:
\be
|R| \simeq 0.01 \frac{\al \om^2}{k_0^2} \, .
\label{highfreq}
\ee
Since the scale of Lorentz breaking is very high, only frequencies of order $\om \sim k_0$ will be reflected appreciably.

The lowest frequency waves emitted by a black hole inspiral have wavelengths comparable to the gravitational radius
of the source, and so in this case $\om r_0 \simeq 1$. It is clear from Eq. (\ref{eq:firstorder}) that in this case all the
contributions to the source ${\cal S}$ are comparable, because $f \simeq 1$. Regulating the integrals at the photon sphere, and accounting for the multiplicity of the various contributing factors, the integral is roughly 
\be
|R| \simeq {\cal O}(1) \frac{\al \om^2}{k_0^2} \simeq  {\cal O}(1) \frac{\al }{k_0^2 r_0^2}  \, .
\ee
The increase of the numerical factor by about two orders of magnitude comes from summing the large coefficients in the expansion. The scaling with $k_0 r_0$ however must follow from the equivalence principle to match the high frequency result (\ref{highfreq}).
The numerical solution for $R$ at all frequencies is shown in Fig.~\ref{fig:R}.

In either case, since most of the observed frequencies of gravity waves are in the range $\omega \sim 1/r_0$, and since $r_0 \sim {\rm km} \sim 10^{9} {\rm eV}^{-1}$, for typical $k_0$ which are high, suppressing Lorentz violations at low energies, the echo signals would be extremely small.
E.g. if we take the scales from particle physics constraints \cite{kostelecky}, $R$ would end up being miniscule. Even if we allow Lorentz violations in the gravitational 
sector at a millimeter scale, $k_0 \ga 10^{-3} {\rm eV}$, the echoes would be feeble: $R\la 10^{-12}$. To get a 
nonegligible signal from Lorentz-violating dispersion relation corrections we would need $k_0 \sim 10^{-9} {\rm eV}$, 
which is excessive. 

For completeness, we note that to determine the time between different echoes in a sequence, we need to estimate the region of origin of 
the incipient reflected wave due to the correction near the horizon. From (\ref{intr}), (\ref{inta}), (\ref{intb}) clearly the strongest contribution to the integral comes from the region where the integrand has an extremum, attaining a stationary phase and adding up coherently to the 
reflection coefficient.  For high frequency waves, where the integral is dominated by (\ref{intb}) this implies
that the region where the reflection occurs is at ${\cal S} \rightarrow 0$, which by Eq. (\ref{Shigh})  means that the 
reflected wave comes from near the horizon, since the wave changes the least in that region\footnote{Using the tortoise coordinate, in this regime the wave grows slowly as $x$ over a very large domain of $x_*$, and picks up contributions of ${\cal O}(\omega^2/k_0^2)$ .}. For low frequency waves, the reflection is more spread out given the
competition between different terms in (\ref{eq:firstorder}). 

Note that if the echoes came from a specific locale deep in the gravitational well of the black hole, the time interval between adjacent reflections, as measured by an observer at a safe distance from the source ($r \gg r_0$) is time-dilated by the black hole's gravitational field. Indeed, the time between two echoes can be estimated by the interval it takes a massless particle to fall from the photon sphere to near the horizon where the echoes originate. This means that the time interval is given by the tortoise coordinate displacement between the echo's origin and the photon sphere,
\be
\Delta t = \Delta r_* \simeq 3r_0/2 - r_{echo} + r_0 \ln\Bigl(\frac{r_0/2}{r_{echo}- r_0}\Bigr) \simeq  r_0 \ln\Bigl(\frac{r_0/2}{r_{echo}- r_0}\Bigr) \, .
\ee
The log in this formula is the time dilation factor. In practice this factor is finite: if the black hole's geometry is modified no closer than $r_{echo} - r_0 \simeq \ell_{Pl}$, this term is 
$\Delta t \la r_0 \ln(r_0/{\ell}_{Pl})$, which for solar mass black holes is about $10^{-3}$ seconds. As noted in \cite{cardoso,kleban}, this makes echoes easy to separate from the primary gravity wave emissions by the black  hole, if they are completely localized. 

In slight contrast, the Lorentz violations which we discussed here would produce a more smeared-up spectrum of echoes which would originate at all distances from the horizon. However most of the power would be stored in the
modes coming from deep inside, meaning that the echo would start as a gradually rising ``hum". Nevertheless, since the power in echoes is so weak, this feature seems moot. 

As we noted in the introduction, the reason why the echoes are quiet is decoupling. The echoes are suppressed because the reflective corrections are perturbative and so very small at low energies. To generate a significant effect, the reflective terms need to mix with waves whose frequencies are comparable to the scale where violations occur.
If the distortions of  GR black holes occur only at short distances, and via perturbative terms, this renders the signals weak, unless some dramatic effects from beyond normal GR and EFT kick in. 

As an example of such effects, it has been speculated in the literature on black hole echoes that a source of dramatic effects might be the firewalls of \cite{polchinski}. Firewalls are purported structures around black holes which  may arise from quantum backreaction of Hawking radiation onto the black hole. Without getting into the details of whether black hole firewalls are really there or not, or what they are, which still seems to be a matter of debate \cite{apologia}, we will merely try to elucidate if they can give rise to observable echoes. Since the firewalls are expected to be layers of energy deviating from the vacuum near the black hole horizon, one might model their distortion of classical black hole geometry by introducing thin shells of matter as  sources of gravitational field in Einstein's equations \cite{hollybolly}, and then matching various bulk solutions across the shell. The constructions of \cite{hollybolly} do this, by matching an exterior Schwarzschild geometry to a section of the interior RN, specifically taking the interior RN to be the section which hosts the timelike singularity, see Fig. 2. 

\begin{figure}[ht]
  \centering
  \includegraphics[width=0.55\textwidth]{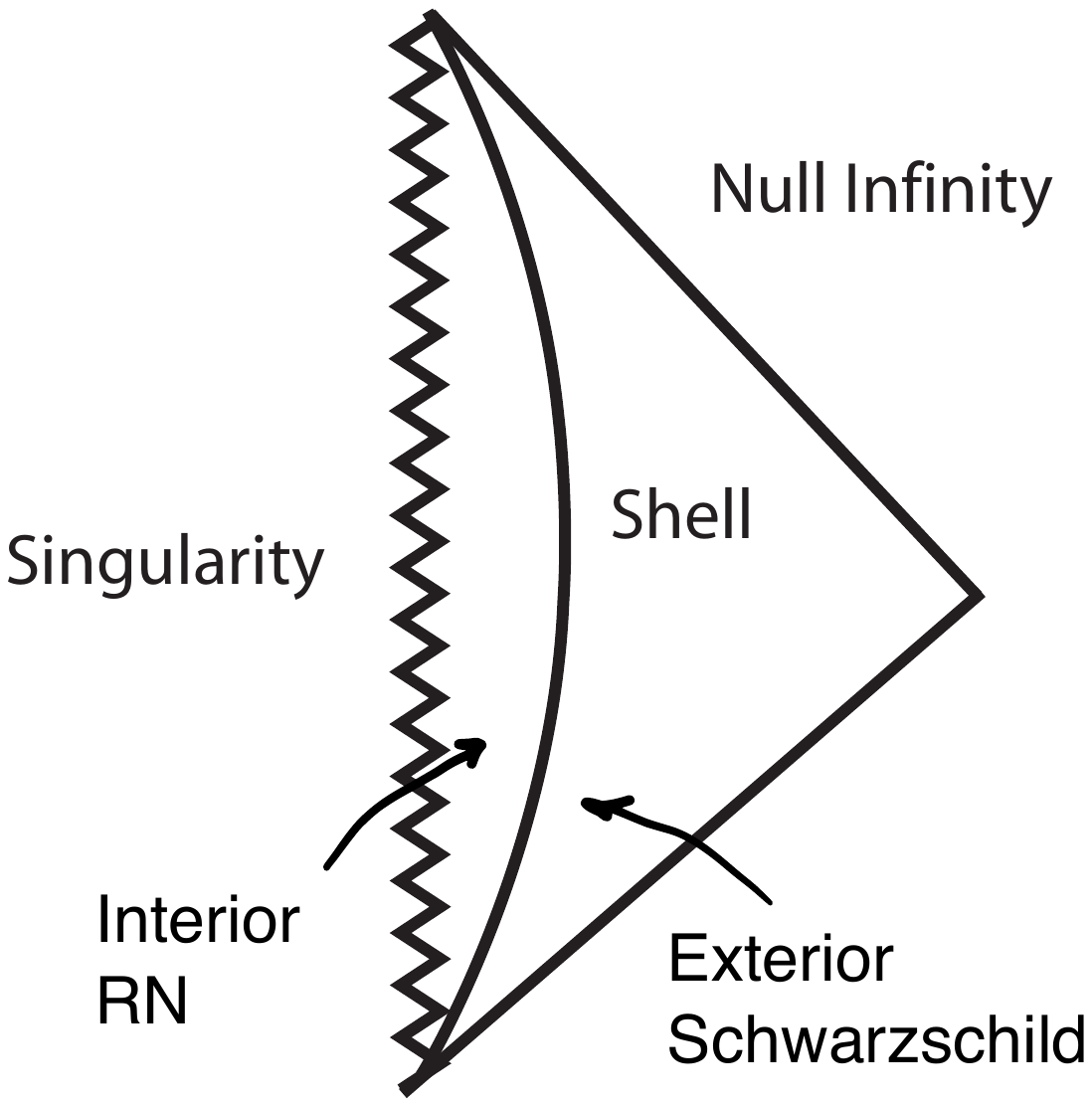}
  \caption{Spacetime diagram of the proposed ``classical firewall'' of \cite{hollybolly} completely excising the horizons. The RN singularity is enclosed by a nearby shell with Planckian density. }
  \label{fig:penrose}
\end{figure}

This configuration can in principle provide significantly stronger echoes. Namely, while the exterior is Schwarzschild, the shell can reflect some of the infalling waves. This is even clear from Huygens' principle. At the level of perturbation
theory description of the wave packets propagating on this configuration, the shell appears as a $\delta$-function correction to the wave equation (\ref{eq:freeeq}), with a coefficient $\alpha \sim \sigma/M^2_{Pl}$, where 
$\sigma$ is the shell tension. The reflection coefficient then is given by
$R \sim \gamma^2 \alpha^2/\omega^2 \sim \gamma^2 
\alpha^2 r_0^2$, where $\gamma$ is the time dilation factor of the clock on  the shell relative to  infinity,
$\gamma  = 1 -r_0/r_s$. A shell placed far from the horizon, and near the photon sphere, would have $\gamma \sim 1$ and $\alpha \sim 1/r_0$, which would make it very reflective, $R\sim 1$. However as noted in \cite{hollybolly}, such a shell is hardly a model of a firewall, and in fact it would make the configuration behave much more like
a star rather than a black hole. On the other hand, a shell placed close to the horizon would have $\alpha \sim M_{Pl}$, but the time dilation factor would be $\gamma \sim \epsilon \sim (\ell_{Pl}/r_0)^2$, so that
$R \sim  (\ell_{Pl}/r_0)^2 \ll 1$. Basically the time dilation would all but extinguish the reflection from the shell itself, due to the covariance of its description. This again displays decoupling. 

However, if the shell completely removes the horizon as in Fig.  2, inside the shell there resides a timelike singularity, which in general need not be absorptive. 
In fact a priori the boundary conditions for a wave equation on the singularity can be quite general, and very reflective. One could, for example, attempt to model the singularity's influence by being guided by decoupling,
as in the singular extra-dimensional bulks in \cite{gellmannzwiebach}. In this case one might
impose the Neumann conditions on the wave function -- which would of course yield maximal reflection. If so, this
could produce a very significant signal. The details are beyond the scope of this article, but we hope to return to
this question elsewhere. 

In summary, in this work we have considered the possibility to generate echoes in the gravity wave sources, due to higher order irrelevant operators that correct the dispersion relations. If the corrections do not alter the black hole
geometry, leaving a regular horizon as a null boundary at the bottom of the black hole's gravitational well, the echoes
will have a miniscule power. The reason is the decoupling between the near horizon high energy physics and the low
energy physics outside of the black hole. We have reflected also on the possibility of firewalls enhancing the signal.
In this case, we used a  recent classical model of a firewall \cite{hollybolly}. In this case, the very correction of the
geometry, parameterized by a shell shielding the horizon, also decouples when the shell is far inside. However,
if the horizon is completely removed, which has been argued in \cite{hollybolly} to be a possibility, leaving a timelike
singularity inside, the echoes may be enhanced dramatically. The singular configuration need not be a perfect absorber across the full spectrum of frequencies. This illustrates that echoes, if observed, are a signature of very dramatic deviations from standard GR or EFT in the vicinity of horizons. It would be interesting to study this example in more detail to explore this exotic but curious possibility.



\vskip.5cm

{\bf Acknowledgments}: 
We would like to thank R. Emparan, M. Meiers and S. Rajendran for useful discussions. NK thanks CERN Theory Division for kind hospitality in the course of this work. NK is supported in part by the DOE Grant DE-SC0009999.


\begin{thebibliography}{99}

\bibitem{LIGO}
B.~Abbott \textit{et al.} [LIGO Scientific and Virgo],
Phys. Rev. Lett. \textbf{116}, no.6, 061102 (2016).

\bibitem{gibbons}
  S.~R.~Das, G.~W.~Gibbons and S.~D.~Mathur,
  Phys.\ Rev.\ Lett.\  {\bf 78}, 417 (1997). 
  
  \bibitem{cardoso}
  V.~Cardoso, L.~C.~B.~Crispino, C.~F.~B.~Macedo, H.~Okawa and P.~Pani,
  Phys.\ Rev.\ D {\bf 90}, no. 4, 044069 (2014); 
  V.~Cardoso, S.~Hopper, C.~F.~B.~Macedo, C.~Palenzuela and P.~Pani,
  Phys.\ Rev.\ D {\bf 94}, no. 8, 084031 (2016); 
  V.~Cardoso and P.~Pani,
  Nat.\ Astron.\  {\bf 1}, no. 9, 586 (2017).
    
  \bibitem{afshordi}
  J.~Abedi, H.~Dykaar and N.~Afshordi,
  Phys.\ Rev.\ D {\bf 96}, no. 8, 082004 (2017); 
  N.~Oshita and N.~Afshordi,
  Phys.\ Rev.\ D {\bf 99}, no. 4, 044002 (2019);
  N.~Oshita, Q.~Wang and N.~Afshordi,
  arXiv:1905.00464 [hep-th];
  Q.~Wang, N.~Oshita and N.~Afshordi,
  Phys.\ Rev.\ D {\bf 101}, no.2,  024031 (2020). 
  

\bibitem{ashton}
G.~Ashton, O.~Birnholtz, M.~Cabero, C.~Capano, T.~Dent, B.~Krishnan, G.~D.~Meadors, A.~B.~Nielsen, A.~Nitz and J.~Westerweck,
[arXiv:1612.05625 [gr-qc]];
J.~Westerweck, A.~Nielsen, O.~Fischer-Birnholtz, M.~Cabero, C.~Capano, T.~Dent, B.~Krishnan, G.~Meadors and A.~H.~Nitz,
Phys. Rev. D \textbf{97}, no.12, 124037 (2018).

  \bibitem{burgess}
  C.~P.~Burgess, R.~Plestid and M.~Rummel,
  JHEP {\bf 1809}, 113 (2018). 
  
  \bibitem{emparan}
  G.~Raposo, P.~Pani and R.~Emparan,
  Phys.\ Rev.\ D {\bf 99}, no. 10, 104050 (2019). 

  \bibitem{foitkleban}
  V.~F.~Foit and M.~Kleban,
  Class. Quant. Grav. \textbf{36}, no.3, 035006 (2019). 

  
  \bibitem{kleban}
  V.~Cardoso, V.~F.~Foit and M.~Kleban,
  JCAP {\bf 1908}, 006 (2019). 

  \bibitem{coates}
A.~Coates, S.~H.~Völkel and K.~D.~Kokkotas,
Phys. Rev. Lett. \textbf{123}, no.17, 171104 (2019).

  \bibitem{glashow}
  S.~R.~Coleman and S.~L.~Glashow,
  Phys.\ Rev.\ D {\bf 59}, 116008 (1999).


\bibitem{kostelecky}
  V.~A.~Kostelecky,
  Phys.\ Rev.\ D {\bf 69}, 105009 (2004).

\bibitem{mattingly}
  D.~Mattingly,
  Living Rev.\ Rel.\  {\bf 8}, 5 (2005).

\bibitem{ellis} 
  J.~Albert {\it et al.} [MAGIC and Other Contributors Collaborations],
  Phys.\ Lett.\ B {\bf 668}, 253  (2008).
  
\bibitem{liberati}
  S.~Liberati,
  Class.\ Quant.\ Grav.\  {\bf 30}, 133001 (2013) .
  
\bibitem{israel}
W.~Israel,
Phys. Rev. \textbf{164}, 1776-1779 (1967). 
  
\bibitem{bekenstein}
J.~D.~Bekenstein,
Phys. Rev. D \textbf{5}, 1239-1246 (1972).

\bibitem{price}
R.~H.~Price,
Phys. Rev. D \textbf{5}, 2419-2438 (1972).
  
\bibitem{jacobson}
  S.~Corley and T.~Jacobson,
  Phys.\ Rev.\ D {\bf 54} (1996) 1568; 
  T.~Jacobson,
  Prog.\ Theor.\ Phys.\ Suppl.\  {\bf 136}, 1 (1999). 
  
    \bibitem{unruh}
  W.~G.~Unruh,
  Phys.\ Rev.\ D {\bf 51}, 2827 (1995).
  
    
  
\bibitem{brand}
  J.~Martin and R.~H.~Brandenberger,
  Phys.\ Rev.\ D {\bf 63}, 123501 (2001); 
  Mod.\ Phys.\ Lett.\ A {\bf 16}, 999 (2001). 

\bibitem{shenker}
  N.~Kaloper, M.~Kleban, A.~E.~Lawrence and S.~Shenker,
  Phys.\ Rev.\ D {\bf 66}, 123510 (2002); 
  N.~Kaloper, M.~Kleban, A.~Lawrence, S.~Shenker and L.~Susskind,
  JHEP {\bf 0211}, 037 (2002). 


  \bibitem{unruhh}
W.~Unruh,
Phys. Rev. D \textbf{14}, 870 (1976).
  
  \bibitem{pata}
A.~Fabbri and J.~Navarro-Salas,
Modeling black hole evaporation, London, UK: Imp. Coll. Pr. (2005) 334 p.
  
  \bibitem{cutoffs}
N.~Kaloper,
Phys. Rev. D \textbf{86}, 104052 (2012).

\bibitem{russrep}
M.~A.~Naimark, ``Linear representations of the Lorentz group", Macmillan (1964).

\bibitem{page}
D.~N.~Page,
Phys. Rev. D \textbf{13}, 198-206 (1976); 
Phys. Rev. D \textbf{14}, 3260-3273 (1976). 



\bibitem{thooft}
  G.~'t Hooft,
  Nucl.\ Phys.\ B {\bf 256}, 727 (1985).


\bibitem{polchinski}
  A.~Almheiri, D.~Marolf, J.~Polchinski and J.~Sully,
  JHEP {\bf 1302}, 062 (2013). 

\bibitem{apologia}
  A.~Almheiri, D.~Marolf, J.~Polchinski, D.~Stanford and J.~Sully,
  JHEP {\bf 1309}, 018 (2013).


\bibitem{hollybolly}
  D.~E.~Kaplan and S.~Rajendran,
  Phys.\ Rev.\ D {\bf 99}, no. 4, 044033 (2019). 

  
    \bibitem{gellmannzwiebach}
  M.~Gell-Mann and B.~Zwiebach,
  Nucl.\ Phys.\ B {\bf 260}, 569 (1985).

\end{thebibliography}
\end{document}